\documentclass{article}
\usepackage[english]{babel}
\usepackage[utf8]{inputenc}
\usepackage{johd}
\usepackage{url}
\usepackage[utf8]{inputenc}
\usepackage{amsmath,amsthm,amsfonts,amssymb,amscd}
\usepackage{lastpage}
\usepackage{enumerate}
\usepackage{fancyhdr}
\usepackage{mathrsfs}
\usepackage{xcolor}
\usepackage{graphicx}
\usepackage{listings}
\usepackage{hyperref}
\usepackage{listings}
\usepackage{float}
\usepackage{tcolorbox}
\usepackage{natbib}
\usepackage{caption}
\usepackage{subcaption}
\usepackage{rotating}
\pdfoutput=1

\title{A Hierarchical Spike-and-Slab Model for Pan-Cancer Survival Using Pan-Omic Data}

\author{Sarah Samorodnitsky$^{a}$, Katherine A. Hoadley$^{b}$, Eric Lock$^{a}$$^{*}$ \\
        \small $^{a}$Division of Biostatistics, University of Minnesota, Minneapolis, USA \\
        \small $^{b}$ Department of Genetics, Lineberger Comprehensive Cancer Center, University of North Carolina at Chapel Hill, \\ \small North Carolina, USA\\
        \small $^{*}$Corresponding author: Eric Lock; \tt{elock@umn.edu} \\
}
\date{}

\begin{document}
\maketitle
\begin{abstract} 
\noindent Pan-omics, pan-cancer analysis has advanced our understanding of the molecular heterogeneity of cancer, expanding what was known from single-cancer or single-omics studies. However, pan-cancer, pan-omics analyses have been limited in their ability to use information from multiple sources of data (%i.e.
e.g., omics platforms) and multiple sample sets (%(i.e. 
e.g., cancer types) to predict important clinical outcomes, like overall survival. We address the issue of prediction across multiple high-dimensional sources of data and multiple sample sets by using exploratory results from BIDIFAC+, a method for
integrative dimension reduction of bidimensionally-linked matrices, in a predictive model. We apply a Bayesian hierarchical model that performs variable selection using spike-and-slab priors which are modified to allow for the borrowing of information across clustered data. This method is used to predict overall patient survival from the Cancer Genome Atlas (TCGA) using data from 29 cancer types and 4 omics sources. Our model selected patterns of variation identified by BIDIFAC+ that differentiate clinical tumor subtypes with markedly different survival outcomes. We also use simulations to evaluate the performance of the modified spike-and-slab prior in terms of its variable selection accuracy and prediction accuracy under different underlying data-generating frameworks. %
Software and code used for our analysis can be found at 
\url{https://github.com/sarahsamorodnitsky/HierarchicalSS_PanCanPanOmics/}.\end{abstract}

\noindent\\

\section{Introduction} 
\label{s:intro}
\subsection{Motivating Application}
Since its completion in 2018, the Cancer Genome Atlas (TCGA) database has become a cornerstone for studying the relationship between cancer molecular heterogeneity and clinical outcomes. TCGA contains data from multiple ``omics" sources, including the genome, transcriptome, proteome, and epigenome, from over 10,000 patients across 33 types of cancer
\citep{hutter2018cancer}, opening the door to pan-omics, pan-cancer research. Pan-omics, pan-cancer research has been motivated by discoveries of vast molecular variation within a single cancer type \citep{cancer2012comprehensive, cancer2014comprehensive, verhaak2010integrated}, as well as discoveries of the same genomic changes affecting tumors from different tissues-of-origin \citep{weinstein2013cancer}. This suggests the importance of considering multiple omics sources and multiple cancer types at once to holistically characterize cancer's etiological landscape. 

One such approach to studying molecular heterogeneity across both omics sources and cancer types is BIDIFAC+, a method of simultaneous factorization and decomposition of variation across bidimensionally linked matrices \citep{lock2020bidimensional}.  BIDIFAC+ identifies latent factors, analogous to principal components, that may be shared across any number of omics platforms or sample sets. These components describe patterns of variability across these combinations of omics sources and sample sets. When applied to TCGA data, BIDIFAC+ revealed patterns of variability shared by mRNA, miRNA, methylation, and protein data driving heterogeneity across multiple cancers \citep{lock2020bidimensional}. However, these results were solely exploratory, and did not consider prediction of important clinical endpoints. Our goal is to assess the prognostic value and clinical relevance of pan-omic patterns of molecular variability identified by BIDIFAC+. To do so, we sought to use a comprehensive model for overall survival that flexibly borrows information across the different types of cancer.

\subsection{Components of our Pan-Cancer, Pan-Omics Analysis} %
%EFL:
Our approach builds on two active areas of statistical methodology: prediction via integrative dimension reduction (Section \ref{bidred}) and structured Bayesian variable selection (Section \ref{spikeandslab}). 

\subsubsection{Prediction via Bidimensional Dimension Reduction}
\label{bidred}
Predictive modeling in the case of a single high-dimensional dataset often begins by first applying a method such as principal components analysis (PCA) to obtain a small set of latent variables (i.e., components) that explain much of the variation in the data 
\citep{massy1965principal}. %EFL: 
These components can be used for predictive modeling using classical approaches \citep{bair2006prediction}.  However, PCA does not translate smoothly to the multi-source (e.g., multi-omics) context. In this context, one may use the results of multi-source integrative methods, like joint and individual variation explained (JIVE) \citep{lock2013joint}, structural learning and integrative decomposition (SLIDE) \citep{gaynanova2019structural}, or generalized integrative principal components analysis (GIPCA) \citep{zhu2020generalized}. %supervised principal components (\cite{bair2006prediction}). 
%EFL:
These methods identify components that are shared across or specific to multiple sources, which  has been shown to improve power and interpretation for multi-omics predictive models over ad-hoc applications of PCA \citep{kaplan2017prediction}.
However, these approaches do not apply when there are multiple sources of covariates and multiple sample sets, as is the case in the pan-omics, pan-cancer setting. This article addresses the issue of prediction across multiple sources of data and multiple sample sets by using components identified by BIDIFAC+ in a predictive model. %EFL:
BIDIFAC+ identifies components that may be shared across any number of sources (e.g., omics platforms) and any number of sample sets (e.g., cancer types).  In particular, we use BIDIFAC+ components from bidimensional integration of multiple omics sources and multiple cancer types to model TCGA patients' overall survival (OS). 

\subsubsection{Bayesian Hierarchical  Spike-and-Slab Survival Model} 
\label{spikeandslab}
In order to model the relationship between patient OS and components from BIDIFAC+ dimension reduction, we consider a Bayesian hierarchical survival regression framework. %EFL:
Bayesian hierarchical regression has been used  previously for pan-cancer survival modeling \citep{samorodnitsky2020pan}, and is attractive in this context because it facilitates borrowing of information across cancer types while allowing a different survival model for each cancer. This feature of our approach is motivated by the assumption that molecular patterns may drive heterogeneity in more than one cancer. However, our model is also flexible enough to allow the effect of these patterns to differ according to the cancer type. Accommodating a censored outcome is straightforward in this framework, which has been demonstrated in prior work \citep{samorodnitsky2020pan,carvalho2010horseshoe}. %In addition, our model includes a variable selection component that is an extension of \cite{george1993variable}'s definition of a spike-and-slab prior, modified to borrow strength across clustered data in determining the model's sparsity structure. 

Many genomic components have little relation to clinical outcomes, and so we pursued a sparse model that accommodates variable selection within the hierarchical framework.  There is an extensive literature on Bayesian approaches to variable selection. \cite{mitchell1988bayesian}, \cite{george1993variable}, and \cite{kuo1998variable} are foundational but differing perspectives on spike-and-slab variable selection. \cite{george1993variable} specify the prior as a mixture between two mean zero Gaussian distributions, one having low variance (the spike) and the other having high variance (the slab). If a coefficient belongs to the slab with high probability under the posterior, this suggests that the corresponding covariate has a non-negligible effect on the response and should be included in the model.  
The spike-and-slab approach is unique in providing an ``included/excluded" interpretation for each predictor through the use of indicator variables that turn on and off each coefficient. In contrast, other Bayesian variable selection approaches focus on adaptively shrinking coefficients of uninformative predictors towards zero. One such example is the Bayesian lasso \citep{park2008bayesian}, which observes that lasso estimates are the posterior mode for coefficients under independent and identically distributed Laplace or double-exponential priors. Another example is the Bayesian elastic net \citep{li2010bayesian}, which formulates a Bayesian model whose posterior mode provides elastic net solutions. The horseshoe prior \citep{carvalho2010horseshoe} is yet another example of a Bayesian shrinkage method that applies a Gaussian prior, with a half-Cauchy prior on its variance term, on the regression coefficients.   

% Redundant paragraph?
%Our model includes a variable selection component that is an extension of \cite{george1993variable}'s definition of a spike-and-slab prior, modified to borrow strength across clustered data in determining the model's sparsity structure. Spike-and-slab variable selection has been proposed in many formulations, but \cite{george1993variable} specify the prior as a mixture between two mean zero Gaussian distributions, one having low variance (the spike) and the other having high variance (the slab). If a coefficient belongs to the slab with high probability under the posterior, this suggests that the corresponding covariate has a non-negligible effect on the response and should be included in the model.  

These Bayesian variable selection methods have been extended in previous work to hierarchical models of many forms. \cite{yang2020consistent} propose using spike-and-slab priors to identify important groups of covariates in nonparametric regression models and seemingly unrelated regressions models. \cite{zhang2014bayesian} propose a variable selection approach which identifies groups of covariates to include in the model and estimates lasso solutions for coefficients in selected groups. These methods operate on a single sample set where inducing sparsity at the group level on covariates is desired. In contrast, \cite{suo2013hierarchical} and \cite{mousavi2014multi} demonstrate the use of spike-and-slab priors for variable selection on a single covariate set shared by multiple sample sets for classification purposes. Hierarchical variable selection has also been considered in Bayesian survival models, as is done in \cite{lee2004bayesian} and \cite{lee2014bayesian}, which both present the use of spike-and-slab priors in proportional hazards models. \cite{maity2020bayesian} considers horseshoe priors that flexibly borrow information across groups in a Bayesian survival model on pan-cancer data.

For our context, we consider a sparse hierarchical model on multiple sample sets. We incorporate multiple covariate sets via the application of BIDIFAC+, reducing the problem to one set of covariates shared across sample sets. To induce sparsity, our model includes a variable selection component which extends \cite{george1993variable}'s definition of a spike-and-slab prior in three ways: (a) we allow the possibility that a predictor is included for one sample group but not another, (b) we allow the slab distribution's location and scale to be inferred hierarchically based on data from groups for which the covariate is included, and (c) we impose a prior on the inclusion probabilities of each covariate to borrow information across groups. 
These modifications adapt the original formulation to borrow information across groups without compromising the flexibility that covariate inclusion and coefficient estimation can differ between groups. This contrasts with \cite{yang2020consistent} and \cite{zhang2014bayesian}, who study variable selection on a single shared covariate set for one sample set, and with \cite{mousavi2014multi} and \cite{suo2013hierarchical}, who study variable selection on a shared covariate set for multiple sample sets but require the same predictors be included for all groups. The work of \cite{maity2020bayesian} most closely resembles ours by considering a setting with multiple sample sets and investigating how to borrow strength while inferring the sparsity structure in the pan-cancer setting. However, using spike-and-slab priors facilitates a natural ``inclusion/exclusion" interpretation not afforded by the horseshoe prior, while still encouraging coefficient shrinkage for selected covariates. Our proposed approach is attractive when it is reasonable to assume groups offer agreeable information about the shared covariates. Then it would be advantageous to borrow strength during variable selection because it increases power to detect which variables are important. Lastly, our model accommodates a potentially censored outcome, often of interest in medical research, so translational models that isolate predictors related to survival are informative. Thus, our proposed model is able to leverage shared information across groups while considering an important clinical outcome. 

The rest of our article is organized as follows. In Section \ref{s:proposed_model}, we state our proposed Bayesian hierarchical model. In Section \ref{s:application}, we apply our proposed model to TCGA data to predict patient OS using patterns of variability identified by BIDIFAC+ and investigate the clinical relevance of the results. In Section \ref{s:simulations}, we present a simulation study evaluating the trade-offs of different approaches to hierarchical variable selection in the context of our data application. Our article concludes with a discussion of the results and suggestions for future work in Section \ref{s:discussion}. 

%\section{Proposed Model} 
\section{Hierarchical spike-and-slab model}
\label{s:proposed_model}
Here we introduce our Bayesian hierarchical model with spike-and-slab priors in general terms, beginning with the classical spike and slab model in Section~\ref{spikeandslab_classic}. We discuss the application of our heirarchical model to dimension reduction results for pan-omic pan-cancer analysis, and survival prediction, in Section~\ref{s:application}. 

\subsection{Spike-and-slab priors}
\label{spikeandslab_classic}

Consider the ordinary linear model for an outcome $y_i$ given covariates $\{X_{i\ell}\}_{\ell=1}^L$,
\[y_{i}=\beta_0+\sum_{\ell=1}^L \beta_{\ell} X_{i\ell}+\epsilon_{i},\]
for subjects $i=1,\hdots,I$.  The classical spike-and-slab model considered by \cite{george1993variable} imposes the following prior on the coefficients $\beta_\ell$: 
\begin{align}
\begin{split}
        \beta_\ell | \gamma_\ell &\sim (1-\gamma_\ell) \hbox{N}(0, \tau_\ell^2) + \gamma_\ell \hbox{N}(0, c_\ell^2 \tau_\ell^2) \\
       \gamma_\ell | \pi_\ell &\sim \hbox{Bernoulli}(\pi_\ell)
\end{split} \label{eq:ssmodel}
\end{align}
where $\tau_\ell^2$ is chosen to be small and $c_\ell^2$ is chosen to be large. 
The indicator $\gamma_\ell$ reflects from which distribution $\beta_\ell$ is generated: if $\gamma_\ell=1$, $\beta_\ell$ is generated from the slab, $\hbox{N}(0, c_\ell^2 \tau_\ell^2)$, and if $\gamma_\ell=0$,  $\beta_\ell$ is generated from the spike, $\hbox{N}(0, \tau_\ell^2)$.  Practically, $\gamma_\ell$ indicates whether covariate $\ell$ has a non-negligible contribution to the predictive model. The prior encourages sparsity and shrinks coefficients under the slab towards zero. Uncertainty in model selection is easy to interpret via the posterior probabilities of each $\gamma_\ell$.    
%Generating $\beta_i$ from the slab reflects signal in the data that suggests covariate $i$ has a non-negligible effect on the response and should be included in the model (\cite{george1993variable}).

\subsection{Hierarchical extensions}

Now, assume the data are grouped or clustered, e.g., by genetic strain or cancer type. We index each group by $i$, $i=1,\dots, I$ and index subjects within each group by $j$, $j=1,\dots, n_i$ where $n_i$ is the sample size for group $i$. Consider $L$ covariates  $\{X_{1}, X_{2}, \dots, X_{L}\}$, where a subset of the $L$ covariates is available for each group. Let $S_i = \{\ell: X_\ell \text{ exists for group } i \}$ be the indices for covariates measured on group $i$. Let $y_{ij}$ be the response for the $j$th subject in the $i$th group, $j = 1, \dots, n_i$, $i = 1, \dots, I$. Specify a linear model for $y_{ij}$ as follows:
\begin{equation}
    y_{ij} = \beta_{i0} + \sum_{\ell \in S_i} \beta_{i\ell} X_{ij\ell} + \epsilon_{ij}
\end{equation}
where $\epsilon_{ij}$ are iid random variables such that $\mathbb{E}(\epsilon_{ij}) = 0$ and $\hbox{Var}(\epsilon_{ij}) = \sigma^2$. It is of note that this framework not only allows for covariate sets to differ between groups, but also allows the effect of each predictor to vary by group, where the partial effect of predictor $\ell$ for group $i$ is given by $\beta_{i\ell}$. We allow for the possibility that a predictor may have no effect on group $i$'s outcome through the use of spike-and-slab variable selection. We extend George and McCulloch's implementation of a spike-and-slab prior \eqref{eq:ssmodel} by inferring the distribution of the slab hierarchically (with a possibly non-zero mean) while allowing for differential inclusion across groups. The hierarchical structure is also extended to the inclusion probabilities. We define our spike-and-slab prior as follows:
\begin{align}
\begin{split}
    \beta_{i\ell}|\tilde\beta_\ell, \lambda^2_\ell, \gamma_{i\ell} &\sim (1-\gamma_{i\ell}) \hbox{Normal}\left(0, z^2\right) + \gamma_{i\ell} \hbox{Normal}(\tilde\beta_\ell, \lambda^2_\ell) \\
    \tilde\beta_\ell &\sim \hbox{Normal}(0, \tau^2) \\
    \lambda^2_\ell &\sim \hbox{Inverse-Gamma}(\alpha_1, \alpha_2) \\
    \gamma_{i\ell}|\pi_\ell &\sim \hbox{Bernoulli}(\pi_\ell) \\
    \pi_\ell &\sim \hbox{Beta}(1,1) 
\end{split}
\end{align}
where $\ell = 1,\dots, L$ and $z^2$ is chosen to be very small. If $\gamma_{i\ell} = 1$, then $\beta_{i\ell}$ is generated from the slab, $\hbox{Normal}(\tilde\beta_\ell, \lambda^2_\ell)$, and if $\gamma_{i\ell} = 0$ then $\beta_{i\ell}$  is generated from the spike, $\hbox{Normal}\left(0, z^2\right)$. Practically, $\beta_{i\ell}$ belongs to the slab if covariate $\ell$ has a non-negligible effect on the response in group $i$.  %its effect on the is generated from the slab if the data from group $i$ suggest covariate $\ell$ has a nonzero effect on group $i$'s response. 
Data from clusters for which covariate $\ell$ is generated from the slab are used to infer the mean $\tilde\beta_\ell$ and variance $\lambda^2_\ell$ of the slab distribution. This may increase our power to infer covariate $\ell$'s effect if the groups provide concordant information. We apply a Beta prior to the inclusion probability $\pi_\ell$ for covariate $\ell$, cementing a fully Bayesian framework. Separate inclusion probabilities for each predictor that are shared across the $I$ groups induces correlation between selected predictors. Consequently, inference on $\pi_\ell$ reflects the proportion of groups for which covariate $\ell$ has predictive power. We implement our model using an in-house Gibbs sampling algorithm. The full conditional distributions for the censored survival model used for our data application in Section~\ref{s:application} are provided in the Appendix.

\section{Application to Pan-Omics, Pan-Cancer Data}
\label{s:application}
We now describe the application of the proposed hierarchical spike-and-slab model to TCGA data to characterize the clinical relevance of components identified by BIDIFAC+ \citep{lock2020bidimensional}. To do so, we model patient OS because it is clearly defined, clinically important, and available for most subjects \citep{liu2018integrated}. The model predictors are derived from applying BIDIFAC+ to TCGA pan-omics, pan-cancer data and are explained in more detail in the following subsection.

\subsection{Dimension Reduction, Data Acquisition, and Cleaning}

 Our data was originally curated for use in \cite{hoadley2018cell} pan-cancer clustering analysis.  These data consisted of 29 cancer types and 4 omics platforms.  The cancer types are primarily defined by their tissue-of-origin, and we denote each type by its TCGA study abbreviation, e.g., BRCA for breast invasive carcinoma and ESCA for esophageal carcinoma.  The omics platforms include (1) RNA-Seq data for 20531 genes, (2) miRNA-Seq data for 743 miRNAs, (3) DNA methylation levels for 22601 CpG sites, and (4) reverse-phase protein array data for 198 proteins. BIDIFAC+ decomposes the data into a sum of low-rank \emph{modules}, each corresponding to structured variation that exists on a subset of the 4 omics platforms and the 29 cancers. Using this method, \cite{lock2020bidimensional} identified 50 low-rank modules from which we derived predictors for our model. 

We obtained predictors from the BIDIFAC+ results by computing the singular value decomposition (SVD) of each low-rank module to identify underlying \emph{components} (analogous to principal components) that are specific to a subset of omics platforms and cancer types. For each module's SVD, we took the product of each singular value with its corresponding right singular vector. This product gives us the component \emph{scores} for each subject, which will serve as predictors in our survival model. In this context, the BIDIFAC+ components are assumed to be independent and roughly orthogonal. Since BIDIFAC+ can produce components that explain negligible variation in the data, we did not want to consider these as possible predictors in our predictive model. We would not expect these components to explain much variability in OS and they would lead to unnecessary noise in our model. To ensure we consider predictors with the highest likelihood of explaining variation in survival, we developed selection criteria that precedes our modeling step. %while retaining components that explained variation in the data. 
Our inclusion criteria were as follows:
\begin{enumerate}
    \item Include the first component from the SVD of each low-rank module. This component explains the most variation within each module. 
    \item Include any other components whose ratio of eigenvalue (squared singular value) to total variability in the original multi-source, multi-cancer data was greater than 0.01. This amounts to selecting predictors that explain at least 1\% of the variation in the original pan-omic, pan-cancer data. 
\end{enumerate}
The 0.01 threshold could be adjusted in future studies but it yielded a manageable number of possible model predictors for our purposes. In sum, we considered 66 components derived from the 50 modules as predictors of OS. We refer to each of these predictors by the module from which it was derived and the index of its corresponding right singular vector from the module's SVD, e.g., predictor 5.1 is the first component from module 5. 

To complete our model, we also included a model intercept for each cancer and patient age at the time of diagnosis as a predictor. In our previous work, we showed that age has a strong effect on overall survival in 27 of the 29 cancers considered here \citep{samorodnitsky2020pan}. We standardized all predictors to have mean $0$ and standard deviation $1$ to facilitate comparisons of covariate effects on survival. 

We obtained clinical data from the TCGA Clinical Data Resource (TCGA-CDR) \citep{liu2018integrated}. Before running any analyses, we removed subjects who were missing both a survival time and a censoring time, removed subjects who had survival times that were negative or zero, and removed subjects missing a value for age. After filtering, we retained 6856 subjects across 29 cancer types with data from the 4 omics platforms. 

\subsection{Model Specification}

We now outline the hierarchical spike-and-slab survival model. Let $y^*_{ij}$ be the (possibly right-censored) event time for the $j$th patient in the $i$th cancer type, $j = 1, \dots, n_i$, $i = 1, \dots, 29$. Then,
\begin{equation}
y^*_{ij} = \begin{cases} y_{ij} \quad \text{if subject is not censored} \\
y_{ij}^c \quad \text{if subject is censored}\end{cases}
\end{equation}
where $y_{ij}^c$ is the censor time for the $j$th subject in the $i$th cancer type. Then we assumed
\begin{equation}
\log y^*_{ij} \sim \mbox{Normal} \left(\beta_{0i} + \sum_{\ell \in S_i} \beta_{i\ell} X_{ij\ell}, \hspace{1mm} \sigma^2 \right)
\end{equation}
where
%\begin{equation}
%X_{ij\ell } = \begin{cases} 
%X_{ij\ell} \quad \text{if covariate $\ell$ available for %$i$th cancer} \\
%0 \quad \text{if covariate $\ell$ not available for %$i$th cancer}
%\end{cases}
%\end{equation}
$S_i = \{ \ell: X_\ell \text{ exists for group } i \} \subseteq \{\text{Age}, 1, 2, \dots, 66\}$ is the set of covariate indices available for group $i$. $\{\text{Age}, 1, 2, \dots, 66\}$ represents the full set of all possible predictors, including patient age and each of the 66 components selected by our filtering step. Due to the fact that not every cancer type was used to construct each BIDIFAC+ module, not all predictors are available for all cancer types. We selected a log-normal likelihood because previous work demonstrated it outperforms other parametric models for pan-cancer survival \citep{samorodnitsky2020pan}.

We now specify prior distributions for each of the model parameters. For $i = 1, \dots, 29$, we used a $\hbox{Normal}(\tilde\beta_0, \lambda^2_0)$ prior for the intercept for each cancer type. The mean and variance $\tilde\beta_0$ and $\lambda^2_0$ are %estimated from the data
inferred hierarchically, with respective priors $\tilde \beta_0 \sim \hbox{N}(0, 10^2)$ and $\lambda^2_0 \sim \hbox{IG}(1, 1)$. These priors were chosen to be sufficiently uninformative and to match the scale of the data. The data application results appeared to be insensitive to the choice of hyperparameters in these priors. The intercept for each cancer type was excluded from the spike-and-slab framework. For cancer $i$, $i=1,\dots, 29$, if covariate $\ell \in S_i$, we used the following hierarchical spike-and-slab prior: 
\begin{align}
\begin{split} 
    \beta_{i \ell}|\gamma_{i\ell} &\sim (1-\gamma_{i\ell}) N\left(0, \frac{1}{10000}\right) + \gamma_{i\ell}N(\tilde \beta_\ell, \lambda^2_\ell) \\
\gamma_{i\ell} | \pi_\ell &\sim \hbox{Bernoulli}(\pi_\ell)
\end{split}
\end{align}
where $\gamma_{i\ell}$ is an inclusion indicator that reflects whether or not a coefficient comes from the spike or the slab distribution. The spike variance was arbitrarily set at $\frac{1}{10000}$ and results were not sensitive to this choice. As described in Section \ref{s:proposed_model}, we infer the mean and variance of the slab distribution for each coefficient based on the data. We used a $\hbox{Normal}(0,1)$ prior for $\tilde\beta_\ell$ and an $\hbox{Inverse-Gamma}(5,1)$ prior for the variance for the effect of each covariate, $\lambda^2_\ell$, again to reflect the scale of the predictors. We used an uninformative prior for the inclusion probability $\pi_\ell$ for each covariate $\ell$, $\pi_\ell \sim \hbox{Beta}(1,1)$. 
%This inclusion probability is shared across all cancer types to allow for borrowing information across cancers while simultaneously allowing differential inclusion of each covariate. 
Lastly, we used an $\hbox{Inverse-Gamma}(0.01, 0.01)$ prior for the shared survival time variance, $\sigma^2$, across cancer types. We fit our model using an in-house Gibbs sampling algorithm and provide the conditional posteriors of each parameter in the Appendix.

\subsection{Model Selection}

We first assessed which of the following model frameworks provided appropriate fit to the TCGA data factorized by BIDIFAC+:

\begin{enumerate}
    \item A hierarchical spike-and-slab model, our proposed model.
    \item A null model, with only a random intercept for each cancer type and no covariates.
    \item A full hierarchical model, with no spike-and-slab component and all covariates included.
    \item A hierarchical model with a spike-and-slab component and prior inclusion probabilities fixed at $0.5$.
    \item A hierarchical spike-and-slab model where a single inclusion probability, $\pi$, is shared for all covariates and all cancer types (as opposed to inferring an inclusion probability, $\pi_\ell$, for each covariate) with a uniform prior $\pi \sim \mbox{Beta}(1,1)$. This model is henceforth referred to as the ``shared" model.
\end{enumerate}
We compared how these models fit the data using 5-fold cross validation of the log-posterior predictive likelihood, which is defined as follows. Let $\vec{Y} = \{\vec{Y}^{\text{train}}, \vec{Y}^{\text{test}} \}$ be the full data split into a training and test set. Let $p(y|\Theta_0, X)$ be the log-normal probability density for survival time, given all model parameters $\Theta_0$ and covariates $X$. On each training fold, we fit the model and generated posterior samples for each parameter. For each posterior sample $t$ after burn-in and thinning, we computed
\begin{equation}
  P(\vec{Y}^{\text{test}}| \Theta_o^t, \vec{X}^{\text{test}}) = \prod_{\substack{(i,j) \\ \text{uncensored}}} p(y_{ij}|\Theta_o^t,X_{ij}) \prod_{\substack{(i,j) \\ \text{censored}}} \Pr(y_{ij} > y_{ij}^c \mid \Theta_o^t, X_{ij})  
\end{equation}
where $\Theta_o^t$ is a vector of all the $t$th iteration posterior samples for the parameters of the probability distribution of survival and $y_{ij}^c$ is the censor time for the $j$th patient in the $i$th cancer type. After computing this quantity for each iteration, we computed an estimate of the out-of-sample posterior predictive likelihood:
\begin{equation}
\int P(\vec{Y}^{\text{test}}|\Theta_0, \vec{X}^{\text{test}}) P(\Theta_0|\vec{Y}^{\text{train}},\vec{X}^{\text{train}})d\Theta_0 \approx 
\frac{1}{T} \sum_{t=1}^T P(\vec{Y}^{\text{test}}|\Theta^t, \vec{X}^{\text{test}})
\end{equation}
where T is the number of sampling iterations after burn-in and thinning. The log-posterior likelihood measures how well a model fits the observed data, with a higher value indicating better fit. After running each model on the training fold and computing the log-posterior predictive likelihood  on the corresponding test fold, we took the average of each models' log-posterior likelihoods to determine which framework provided the best fit. 

\subsection{Data Application Results}
Our model selection results are as follows: the proposed hierarchical spike-and-slab model had a mean out-of-sample log-posterior likelihood of -1018.803, followed by the shared model with a posterior likelihood of -1022.285, then the model with prior inclusion probability fixed at 0.5 (-1034.535), the null model (-1051.539), and the full hierarchical model without a spike-and-slab component (-1058.665). Based on these results, the proposed hierarchical spike-and-slab model provided the best fit for overall patient survival so we proceeded with it for the rest of our analysis. 

\iffalse
\begin{table}[ht]
\centering
\begin{tabular}{c|c}
\hline 
Model Type & Mean Log-Posterior Likelihood \\
\hline
Hierarchical Spike-and-Slab & -1018.803 \\
Prior Inclusion Prior Shared Across Covariates & -1022.285 \\
Prior Inclusion Fixed at 0.5 & -1034.535 \\
Null Model & -1051.539 \\
Hierarchical Without Spike-and-Slab & -1058.665 
\hline
\end{tabular}
\vspace{10mm}
\caption{Model selection results using 5-fold cross validation of the log-posterior likelihood}
\label{tab:model_comparison_results}
\end{table}
\fi

\begin{sidewaysfigure}
    \centering
    \includegraphics[scale=0.32]{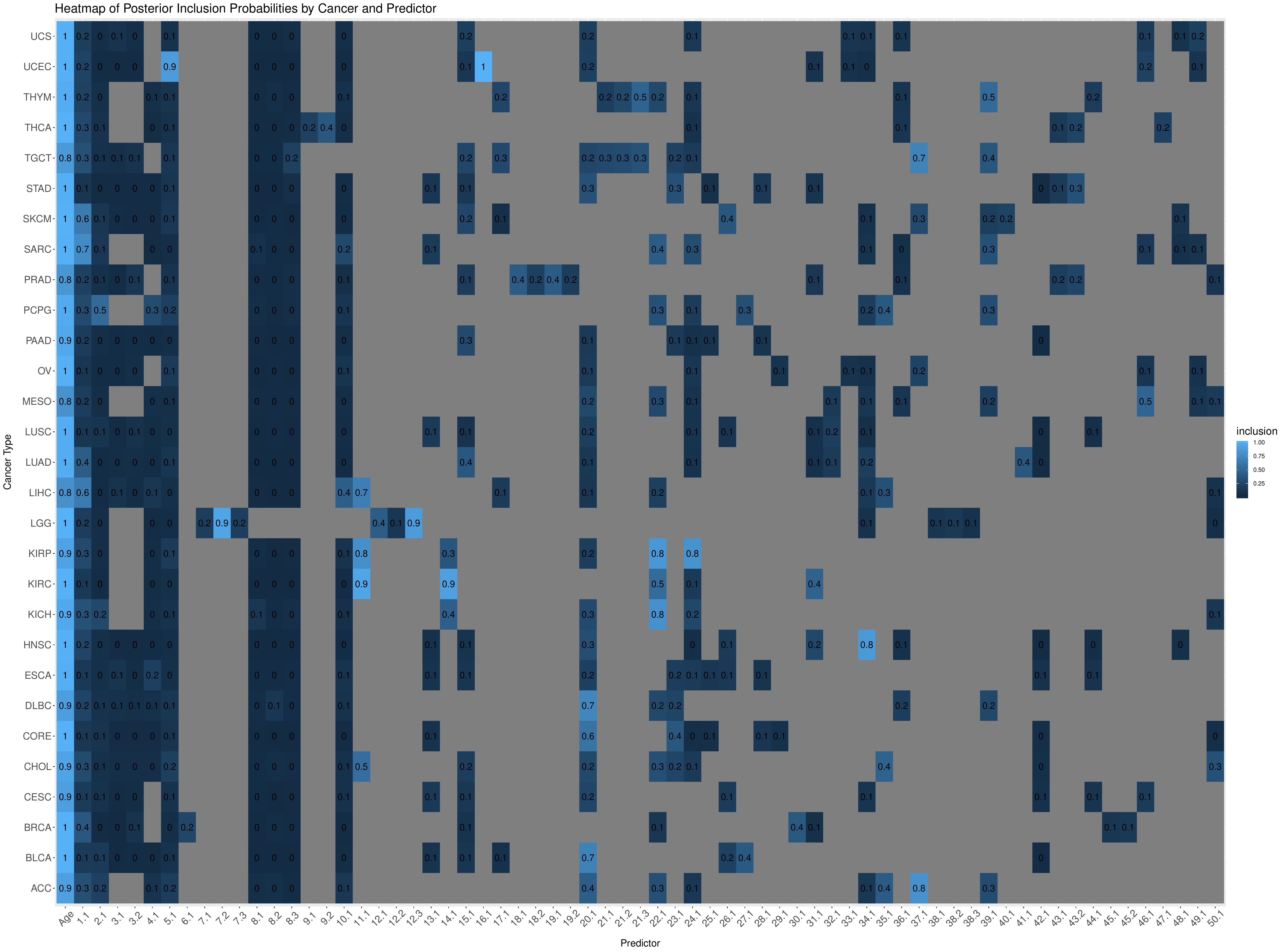}
    \caption{Posterior inclusion probability heatmap for every cancer type and every predictor. The value printed on each box is the posterior probability of inclusion. Gray space indicates a predictor was not available for a particular cancer type. Brighter blue colors indicate higher probability of inclusion, while deeper blue indicators indicate lower probability of inclusion.}
    \label{fig:heatmap}
\end{sidewaysfigure}

We ran the hierarchical spike-and-slab model on the factorized TCGA data for 100000 iterations with a 50000 iteration burn-in and 10-iteration thinning.  Multiple runs of the model with different initial values gave similar results, suggesting that convergence was satisfactory.  We display the variable selection results in Figure \ref{fig:heatmap} via a heatmap of the posterior inclusion probabilities. The posterior inclusion probability for covariate $\ell$ in cancer $i$ is the average of its inclusion indicators generated by our model after burn-in and thinning. 
%The squares in this heatmap range from dark blue (low probability of inclusion) to bright blue (high probability of inclusion) with grey squares blocking out covariates that were not available for particular cancer types.  
Age was included for every cancer type with uniformly high probability, while the inclusion of pan-omic components were comparatively sparse. BIDIFAC+ predictors that capture molecular variation across all or most cancer types were mostly not included by our model. However, certain BIDIFAC+ predictors were identified as predictive of patient survival with high probability and are summarized in Table \ref{tab:inclusion_results}, ordered by descending posterior inclusion probability.
%(highlighted in the ``Probability" column). 
In total, our hierarchical spike-and-slab model selected 24 BIDIFAC+ components across 17 cancer types, based on a posterior inclusion probability above 0.5. 
%Table \ref{tab:inclusion_results} also provides an estimate of the mean effect of that covariate on survival and an interval estimate for its effect. 

%\newpage

Note that the sign of the effects and credible intervals are not immediately interpretable because the identified components (given by singular vectors of an SVD) are uniquely defined up to their sign. However, we can interpret the scale of the effect. Motivated by these results, we chose to investigate more deeply components included for uterine corpus endometrial carcinoma (UCEC), brain lower grade glioma (LGG), kidney renal papillary cell carcinoma (KIRP), kidney renal clear cell carcinoma (KIRC), and kidney chromophobe (KICH) to understand the clinical relevance of the pan-omic components selected. In what follows, we describe our investigation into the clinical significance of these components.  While the figures and discussion describe the marginal effects of components, bear in mind that because our model is multivariate the identification of a component's predictive power for survival is relative to the information contained in other model predictors.  We describe here our investigation into components 16.1 for UCEC, 7.2 for LGG, and 11.1 for KIRP and KIRC; we  explore the inclusion of additional components for UCEC, LGG, and KICH in Section 1 of the supplementary material of this article. 

\begin{figure}
\begin{subfigure}{.5\textwidth}
\centering
\includegraphics[width=\linewidth]{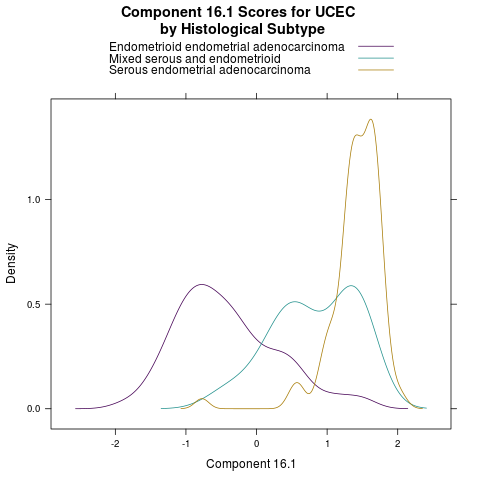}
\caption{  }
\label{fig:ucec_a}
\end{subfigure}%
\begin{subfigure}{.5\textwidth}
  \centering
  \includegraphics[width=\linewidth]{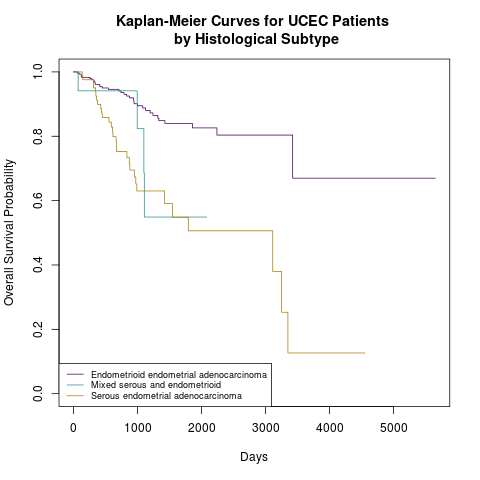}
  \caption{  }
  \label{fig:ucec_b}
\end{subfigure}
\begin{center}
\end{center}
\caption{Figure \ref{fig:ucec_a} displays a KDE plot for the selected component 16.1, which was identified as predictive of survival in UCEC subjects. Component 16.1 scores for subjects with serous UCEC cluster separately from subjects with endometrioid and mixed UCEC. Figure \ref{fig:ucec_b} shows the Kaplan-Meier survival curves for each of the histological subtypes, with the serous subtype showing the worst survival.}
\label{fig:UCECPlots}
\end{figure}

BIDIFAC+ component 16.1 was identified as predictive of survival by the model with near certainty (Table \ref{tab:inclusion_results}). We investigated if this component was associated with UCEC's three histological subtypes: endometrioid, serous, and mixed serous and endometroid \citep{levine2013integrated}. We examined this using histological labels provided in TCGA-CDR \citep{liu2018integrated}. Based on the kernel density estimation (KDE) graph shown in Figure \ref{fig:ucec_a}, the three UCEC histological subtypes cluster distinctly along component 16.1. This suggests that this pattern of variation is primarily driven by distinctions between the three types of UCEC tumors. The Kaplan-Meier survival figure provided in %figure 
Figure \ref{fig:ucec_b} shows divergent survival outcomes for the three subtypes. \cite{levine2013integrated} found that serous and serous-like tumors show extensive somatic copy number alterations (SCNAs), while endometrioid tumors do to a lesser degree, and observed that SCNAs roughly correlated with progression-free survival. While we modeled OS, this may be an underlying latent variable.

% UCEC 1x2 Figure

\begin{figure}
\begin{subfigure}{.5\textwidth}
\centering
\includegraphics[width=\linewidth]{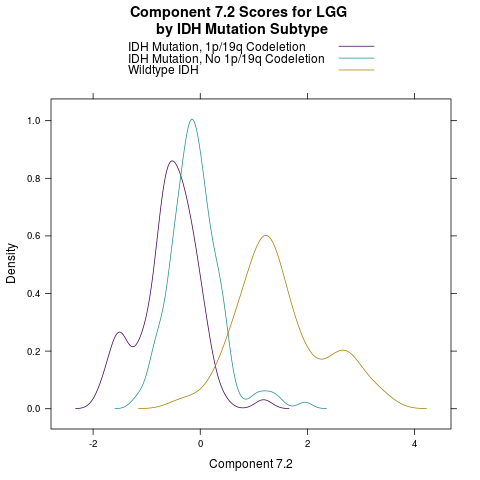}
\caption{  }
\label{fig:lgg_a}
\end{subfigure}%
\begin{subfigure}{.5\textwidth}
  \centering
  \includegraphics[width=\linewidth]{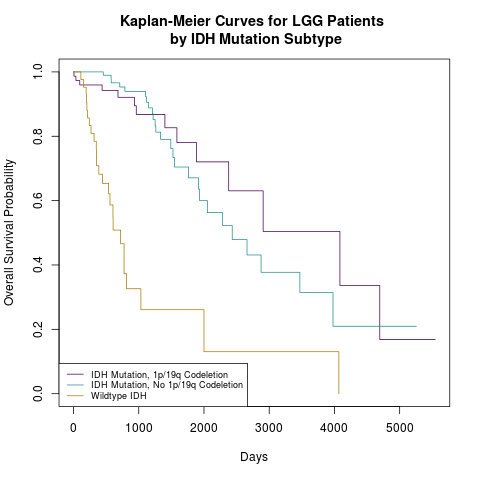}
  \caption{  }
  \label{fig:lgg_b}
\end{subfigure}
\begin{center}
\end{center}
\caption{Figure \ref{fig:lgg_a} displays a KDE plot for the selected component 7.2, which was identified as predictive of survival in LGG patients. Estimated distributions are colored by IDH mutation and 1p/19q codeletion status.
Figure \ref{fig:lgg_b} shows the Kaplan-Meier survival curves for each of the mutation subgroups, with the IDH wildtype mutation showing worst overall survival.}
\label{fig:LGGPlots}
\end{figure}

BIDIFAC+ component 7.2 was identified as predictive of survival in LGG subjects with a posterior probability of $0.92$. We considered its association with the mutation status of genes IDH1 and IDH2 and deletion status in chromosome arms 1p and 19q (1p/19q codeletion) using data from \cite{cancer2015comprehensive}. Mutations in these genes define most cases of LGG and contribute to an LGG subtype associated with better survival \citep{cancer2015comprehensive}. We saw subjects with wild-type IDH mutation clustered distinctly along component 7.2 (Figure \ref{fig:lgg_a}), with Kaplan-Meier survival curves in Figure \ref{fig:lgg_b} displaying divergent survival patterns among the three IDH mutation groups. This suggests that this pattern of variation is linked to IDH mutation patterns that correlate with patient survival. 

% LGG 1x2 Figure

%which have been shown to be highly associated with glioma-CpH Island Methylator Phenotype (\cite{noushmehr2010identification}).

\begin{figure}
\begin{subfigure}{.5\textwidth}
\centering
\includegraphics[width=\linewidth]{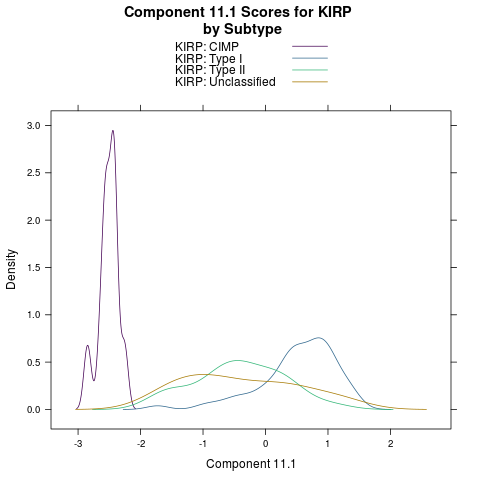}
\caption{  }
\label{fig:kirp_a}
\end{subfigure}%
\begin{subfigure}{.5\textwidth}
  \centering
  \includegraphics[width=\linewidth]{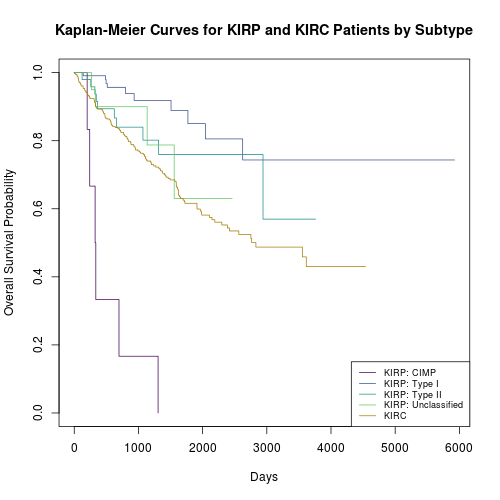}
  \caption{  }
  \label{fig:kirp_b}
\end{subfigure}
\begin{center}
\end{center}
\caption{Figure \ref{fig:kirp_a} displays a KDE plot for component 11.1 within each of the KIRP subtypes, showing CIMP subjects clustering distinctly along component 11.1. This pattern of variation was identified as predictive of survival in KIRP as well as KIRC subjects. Figure \ref{fig:kirp_b} shows a Kaplan-Meier survival plot for all KIRP and KIRC subjects, showing KIRP subjects with the CIMP subtype have the poorest survival. Though KIRC does not currently have a known CIMP subtype, the clinical significance of this subtype is well-known for KIRP. Our analysis suggests a similar distinction may exist in KIRC that is also clinically relevant.   }
\label{fig:KIRPPlots}
\end{figure}

Lastly, component 11.1 was associated with survival for KIRP and KIRC subjects. Using the classification scheme from TCGA's pan-renal project, samples were documented as either KIRP or KIRC, with KIRP further subdivided into type I, type II, and CIMP. Any KIRP patients who did not fit into these categories were left unclassified. CIMP refers to a CpG island methylator phenotype \citep{ricketts2018cancer} and type I and type II are characterized by specific genetic mutations \citep{cancer2016comprehensive}. The CIMP subgroup is known to have the poorest survival of all renal cancers \citep{ricketts2018cancer}, which prompted us to examine if this pattern of variation captured this distinction using data from \cite{ricketts2018cancer}. Figure \ref{fig:kirp_a} shows KIRP subjects classified as CIMP cluster separately from the other three subgroups. Kaplan-Meier survival curves emphasize the stark survival difference between CIMP and the remaining KIRP and KIRC subjects. KIRC also shows poorer survival compared to KIRP types I, II, and unclassified subjects (Figure \ref{fig:kirp_b}). While it seems that 11.1 is associated with KIRP clinical subtypes (e.g., CIMP), it is unclear to what characteristics of KIRC these predictors are linked. The presence and clinical relevance of the CIMP phenotype has been well-studied for KIRP, and our analysis suggests that similar distinctions exist within KIRC that are also clinically relevant.

% KIRP KIRC 1x2 Figure

\section{Simulations}
\label{s:simulations}
We now present a simulation study to compare different approaches to hierarchical variable selection. The goal of our simulation study was to characterize how modifications to the hierarchical variable selection component of our proposed model perform under different data-generating schemes, specifically in the context of our data application. We achieved this by comparing our model under various data-generating schemes to other models specifically tailored to match the way the data was generated. 

We designed the data-generating schemes to mimic our TCGA data application in Section \ref{s:application} by generating groups of the same sample size, with the same number of covariates for each group, and by randomly right-censoring subjects. The degree of overlap across groups for each covariate matches that in our data application, which offers the flexibility that some covariates are shared across all groups, some covariates are shared across subsets of groups, and some covariates are present on only one group.  %Allowing groups to have their own unique predictors reflects the possibility that future investigators may want to consider clinical variables like stage and grade which are uniquely defined for each cancer. 
Each model assumed a log-normal outcome and approximately 50\% of subjects were censored. We considered the following five modeling frameworks:
\begin{enumerate}
    \item A hierarchical spike-and-slab model, our proposed model.
    \item The shared model defined in the ``Model Comparison" subsection of our data application in Section \ref{s:application}.
    \item A model in which the inclusion probabilities are fixed at 0.5.
    \item A hierarchical model with no spike-and-slab component, so all covariates are included.
    \item An intercept-only model that excludes all covariates. 
\end{enumerate}
We considered six data-generating scenarios:
\begin{enumerate}
    \item Each covariate is included for all groups for which it is available with probability 0.5 or excluded for all groups for which it is available with probability 0.5.
    \item  Each covariate is included for all groups for which it is available with probability 0.1 or excluded for all groups for which it is available with probability 0.9.
    \item Each covariate is included independently for each group with probability 0.5, i.e. no true hierarchical structure.
    \item Each covariate is included independently for each group with probability 0.1.
    \item All covariates are included in the model.
    \item All covariates are excluded in the model.
\end{enumerate}
We compared the performance of each model using two metrics: the mean sum-of-squared deviations (SSD) between the true inclusion indicator and the posterior inclusion estimated by each model and the log-posterior predictive likelihood we defined in Section \ref{s:application}. The mean SSD provides a measure of selection accuracy while the log-posterior predictive likelihood provides a measure of predictive accuracy. 

The mean sum-of-squared deviations is defined as follows. Assume the true inclusion indicator for covariates $\ell$ available for group $i$ to be $\gamma_{i\ell}$. %$\phi_{i\ell}$, $i=1,\dots, 29$ and $\ell \in S_i$. 
Let $\hat\gamma_{i\ell}$ be the posterior inclusion probability for the covariates of group $i$ estimated by the model; $\hat\gamma_{i\ell}$ is computed by averaging the inclusion indicators from each model iteration after burn-in and thinning to ease computational burden. The mean SSD for model $k$, $k=1,\dots, 5$ is
\begin{equation}
  SSD_k = \frac{1}{M}\sum_{i=1}^{29} \sum_{l \in S_i} (\gamma_{i \ell} - \hat\gamma_{i\ell})^2  
\end{equation}
where $M = \sum_{i=1}^{29} |S_i|$ is the total number of regression coefficients in the model. The mean SSD measures the accuracy of a model in selecting covariates that are related to the response with a lower mean SSD reflecting more accurate performance. 

We also compared our models based on their log-posterior predictive likelihoods. To do this, we generated a training data set under the corresponding data-generating condition. After fitting the model on this training data set, we computed the log-posterior likelihood on a test data set which is generated under the same condition and with the same true parameter values as the training data set. We averaged the resulting log-posterior likelihoods and the resulting mean SSDs. 

We designed our simulation as follows. For 20 replications,

\begin{enumerate}
    \item Run each of the five considered models under the six data-generating conditions.
    \item For each condition, generate the data accordingly and run each model for 10000 iterations.
    \item After a 5000 iteration burn-in and 10-iteration thinning, compute the mean sum-of-squared deviations, averaged over the total number of regression coefficients. 
    \item Next, generate a test data set under the same data-generating conditions. Using the posteriors samples generated based on the training data in step 3, compute the log-posterior predictive likelihood. 
    \item At the end of the simulation, average the 20 replications.
\end{enumerate}

We used 20 replications to ensure consistent results when the simulation study was repeated. The resulting mean SSDs and log-posterior likelihoods are shown in the Tables \ref{tab:SSDTable} and \ref{tab:PLTable}, respectively, where we bold the best performing model. In both tables, we use pairwise t-tests as a simple way to assess
whether the observed differences in performance across the simulation replications are statistically significant. If the performance of two models were not significantly different under a particular condition, they were considered to have performed equally well. 

The proposed hierarchical spike-and-slab model performs best under conditions (1) and (2) because each group affords concordant information about each predictor. In this case, it is beneficial to borrow strength across groups when estimating the prior inclusion probability. Unlike conditions (1) and (2), conditions (3) through (6) are not specifically suited to the proposed model but comparable mean SSDs and posterior likelihoods demonstrate its competitive performance. Under condition (3), when the prior inclusion probability was 0.5 for all covariates, the fixed-at-0.5 model naturally performs best. The shared model performs well under both (3) and (4) because there was no added benefit to estimating the prior inclusion probability separately for each covariate. Conditions (5) and (6) were the most extreme, under which all or none of the covariates were included, respectively. Under (5) and (6), the full model (a hierarchical model without a spike-and-slab component) and the null model perform best, respectively. Like (3) and (4), the shared model is competitive because there was no added advantage to estimating a prior inclusion probability for each covariate; however, our proposed model remained comparable in selection and prediction accuracy. 

This simulation study demonstrates that the proposed hierarchical spike-and-slab model is competitive in its ability to correctly identify which covariates to include under the data-generating scenarios considered here. While it offers flexibility to estimate different inclusion probabilities for each covariate and pool information across groups, it can also perform well when borrowing strength isn't advantageous. Under each condition, the proposed model fit the test data well, exhibiting its strength in identifying a model with predictive power. While it may not be the optimal model under each condition considered here, its performance was consistent with models that were specifically tailored to perform well, making it a flexible option for an array of underlying data-generating mechanisms. 

\section{Discussion}
\label{s:discussion}
In this article, we propose a Bayesian hierarchical model with spike-and-slab priors that borrow information across groups in determining the model's sparsity structure. This is achieved by allowing the inclusion probability for a covariate to be shared across groups. This spike-and-slab framework has the advantage of increasing power to detect covariate inclusion and covariate effect while being flexible enough to allow each group to have a different covariate set. It also induces correlation under the posterior between selected predictors. An additional perk of this prior is the natural ``inclusion/exclusion" interpretation that other shrinkage methods, like the horseshoe prior \citep{carvalho2010horseshoe} or the Bayesian lasso \citep{park2008bayesian}, lack. 

We apply this model to TCGA data where we use patterns of variability derived from BIDIFAC+ integration of pan-omics, pan-cancer data as predictors in a model for overall patient survival. Predictive modeling using these data expands upon the exploratory work of \cite{lock2020bidimensional} and contributes to the body of research in prediction using multi-source, multi-sample set data. Factorizing the pan-omic, pan-cancer data prior to predictive modeling is useful because the original genomic data is very high dimensional which presents issues of multicollinearity for modeling. Our model gave sparse results regarding selected predictors that explain variability across a large number of cancers. However, it did identify several patterns within smaller subsets of cancer types that are strongly informative of survival, including clinically relevant molecular distinctions that have been previously established (e.g., subtypes within UCEC and LGG) and similar effects across cancer types that warrant further investigation (e.g., for the kidney cancers). In our context, we assumed BIDIFAC+ components were independent and orthogonal; however, the method could be extended to incorporate correlation in the components if a correlation structure is known a priori. Other worthwhile future directions include considering different parametric assumptions for the survival model and relaxing the assumption that the error variance is shared across groups. Additionally one could consider non-linear models and generalized additive models in this context. Including other clinical covariates, like stage and grade, might elucidate the effect of BIDIFAC+ predictors on overall survival; however, stage and grade are not uniformly defined over different cancer types, which presents a challenge for their use in pan-cancer clinical modeling. Alternatively, one may consider other clinical variables as the response, like progression-free survival, but the availability of such data is not as widespread for the TCGA cohort \citep{liu2018integrated}. Patterns of variation identified by BIDIFAC+ on other omics sources, such as copy number variation, could also be considered as predictors. 

We present results from a simulation study where we evaluate the performance of modifications to the variable selection component in our proposed Bayesian model, specifically in our data application context. The goal of this study was to characterize the flexibility of our proposed model in fitting a diverse array of data-generating schemes that mimic our application's group structure. Our simulation study showed that the proposed model was competitive under all six data-generating conditions considered. This simulation study could be expanded to compare the proposed model to other survival models, such as the proportional hazards model or models assuming different parametric survival distributions. Incorporating other Bayesian variable selection methods, like the horseshoe prior \citep{carvalho2010horseshoe} and the Bayesian lasso \citep{park2008bayesian}, into these models would also be worthwhile for comparison. 

In general, more work can be devoted to devising these variable selection methods to borrow information across grouped data. It would be valuable to evaluate and compare the performance of these extensions, in addition to the spike-and-slab model we discuss here, to characterize their relative advantages and disadvantages in a hierarchical setting.

\section{Software}
\label{sec5}

R code, data, and complete documentation is available at our GitHub respository:
\url{https://github.com/sarahsamorodnitsky/HierarchicalSS_PanCanPanOmics/}.

\section*{Acknowledgments}
This work was supported by the National Institutes of Health (NIH) National
Cancer Institute (NCI) grant R21CA231214-01.

{\it Conflict of Interest}: None declared.

\appendix

\section{Model Fitting Algorithm Details}
\label{s:appendix}
We used an in-house Gibbs sampler to estimate the parameters of our model. At each iteration of the algorithm, we generated a sample of each parameter from its respective conditional posterior distribution. In this section, we outline the conditional posterior of each parameter in our model. 

The conditional posterior of $\pi_\ell$ for the $\ell$th covariate is:
\[
\pi_\ell | \gamma_{.\ell} = \hbox{Beta}\left(1 + \sum_{j=1}^{T_\ell} \gamma_{j\ell}, \hspace{2mm} 1 + T_\ell -\sum_{j=1}^{T_\ell} \gamma_{j\ell}\right)
\]
where $\gamma_{.\ell}$ represents the vector of inclusion indicators for the $\ell$th covariate for all 29 cancer types and $T_\ell$ is the number of cancer types that for which the $\ell$th covariate is available. 

The conditional posterior for $\gamma_{i\ell}$, the inclusion indicator for the $i$th cancer type and $\ell$th covariate is:
\[
P(\gamma_{i\ell} =1 \mid \beta_{i\ell}, \tilde\beta_\ell, \lambda^2_\ell)  = \frac{\pi_\ell \hbox{N}(\beta_{i\ell}; \tilde\beta_\ell, \lambda^2_\ell)}{\pi_\ell \hbox{N}(\beta_{i\ell};\tilde\beta_\ell, \lambda^2_\ell) + (1-\pi_\ell) N(\beta_{i\ell};0,\frac{1}{10000})}
\]

where $N(\cdot, \cdot)$ refers to the density of a normal distribution. \\

The vector of coefficients for the $i$th cancer type, $\beta_{i.}$ has the following conditional posterior:
\[
\beta_{i.} \mid \{X_i, Y_i, \gamma_{i.}, \sigma^2, \lambda^2, \tilde\beta \} \sim \hbox{Normal}( Bb, B)
\]
where 
\[
B = \left[ \frac{1}{\sigma^2} X_i^T X_i + \Sigma_i^{-1}   \right]^{-1}
\]
and
\[
b = \frac{1}{\sigma^2} X_i^T y_i + \Sigma^{-1} diag(\gamma_{i.}) \tilde\beta
\]

based on the results of \cite{Lindley1972Linear}. Here, $X_i$ represents the covariate set for group $i$. $y_i$ represents the outcome vector for group $i$. 

The conditional posterior for $\tilde\beta_\ell$ is:
\[
\tilde\beta_\ell \mid \{ \beta_{.\ell}, \lambda^2_\ell \} \sim \mbox{Normal}\left(\frac{K_\ell \tau^2 \Bar{\beta}_\ell}{\lambda_\ell^2 + K_\ell\tau^2}, \frac{\lambda_\ell^2\tau^2}{\lambda_\ell^2 + K_\ell \tau^2}\right)
\]
where $K_0 = 29$ because the model for every cancer type has an intercept, $K_\ell$ for $\ell = 1, \dots, 67$ is the number of cancer types that have this covariate and are not in the spike, and $\Bar{\beta}_\ell = \frac{1}{K_\ell}\sum_{i=1}^{K_\ell} \beta_{i\ell} $. $\tau^2$ is fixed at $10^2$ for $\ell = 0$ and $1$ for $\ell = 1, \dots, 67$.  

The conditional posterior for $\lambda^2_0$ is:
\[
\lambda^2_0 \mid \{ \beta_{.0} , \tilde\beta_{0} \} \sim \mbox{Inverse-Gamma} \left( \frac{K_0}{2} + 1, 1 + 0.5 W_0  \right)
\]
where $K_0=29$ because the model for every cancer type has an intercept. $W_0 = \sum_{i=1}^{K_0} (\beta_{i0} - \tilde\beta_{0})^2$. 

The conditional posterior for $\lambda^2_\ell$, for $\ell = 1, \dots, 67$ is:
\[
\lambda^2_\ell \mid \{ \beta_{.\ell} , \tilde\beta_{\ell} \} \sim \mbox{Inverse-Gamma} \left( \frac{K_\ell}{2} + 5, 1 + 0.5 W_\ell  \right)
\]
where $K_\ell$ is the number of cancer types for which this covariate is available and also is not in the spike. $W_\ell = \sum_{i=1}^{K_\ell} (\beta_{i\ell} - \tilde\beta_{\ell})^2$. $W_\ell$ only includes the $\beta_{i\ell}$ which are not in the spike. 
The conditional posterior for $\sigma^2$ is:
\[
    \sigma^2 \mid \{ X, Y\} \sim \mbox{Inverse-Gamma}\left(\frac{N}{2} + 0.01, \frac{1}{2}B + 0.01\right)
\]
where $N$ is the number of observations in the model and $B = \sum_{i=1}^{29} (y_i - X_i \beta_{i.})^2$.

\bibliographystyle{plainnat}
\bibliography{bibliography.bib}

\begin{table}[H]
\caption{Variable selection results from hierarchical spike-and-slab model. The ``Component" column gives the module and component number that was selected, the ``Cancer" column gives the cancer for which it was selected, ``Mean Effect" gives the mean posterior draw for the coefficient of the selected covariate, ``Posterior Inclusion Probability" gives the average of inclusion indicators, and ``Credible Interval" gives 95\% credible interval for coefficient effect.}
\label{tab:inclusion_results}
\centering
\begin{tabular}{cccccc}
  \hline
 & Component & Cancer & Mean Effect &  Credible Interval & Posterior Inclusion Probability \\ 
  \hline

  1 & 16.1 & UCEC & -0.50 & (-0.723, -0.273) & 1.00 \\ 
  2 & 11.1 & KIRC & 0.36 & (-0.004, 0.678)  & 0.92 \\ 
  3 & 7.2 & LGG & -0.47& (-0.773, 0.004) & 0.92 \\ 
  4 & 14.1 & KIRC & -0.30 & (-0.59, 0.005) & 0.88 \\ 
  5 & 12.3 & LGG & -0.29 & (-0.551, 0.008) & 0.86 \\ 
  6 & 5.1 & UCEC & -0.35 & (-0.62, 0.006) & 0.86  \\ 
  7 & 34.1 & HNSC & -0.33 & (-0.599, 0.007) & 0.84 \\ 
  8 & 22.1 & KIRP & 0.77 & (-0.011, 1.435) & 0.82 \\ 
  9 & 24.1 & KIRP & -0.62 & (-1.218, 0.012) & 0.79 \\ 
  10 & 37.1 & ACC & 0.54 & (-0.013, 1.228) & 0.79 \\ 
  11 & 22.1 & KICH & 0.53 & (-0.012, 1.191) & 0.78 \\ 
  12 & 11.1 & KIRP & 0.36 & (-0.016, 0.964) & 0.77 \\ 
  13 & 1.1 & SARC & 0.30 & (-0.012, 0.662) & 0.74 \\ 
  14 & 20.1 & DLBC & 0.54 & (-0.017, 1.628) & 0.70 \\ 
  15 & 37.1 & TGCT & 0.49 & (-0.017, 1.436) & 0.68 \\ 
  16 & 20.1 & BLCA & -0.21 & (-0.516, 0.012) & 0.68 \\ 
  17 & 11.1 & LIHC & 0.21 & (-0.079, 0.646) & 0.66 \\ 
  18 & 1.1 & LIHC & 0.25 & (-0.014, 0.659) & 0.63 \\ 
  19 & 1.1 & SKCM & 0.18 & (-0.014, 0.473) & 0.63 \\ 
  20 & 20.1 & CORE & -0.15 & (-0.434, 0.015) & 0.57 \\ 
  21 & 11.1 & CHOL & 0.09 & (-0.34, 0.673) & 0.54 \\ 
  22 & 22.1 & KIRC & 0.12 & (-0.015, 0.452) & 0.51 \\ 
  23 & 39.1 & THYM & -0.28 & (-1.052, 0.018) & 0.51 \\ 
  24 & 2.1 & PCPG & 0.45 & (-0.016, 1.642) & 0.50 \\ 
   \hline
\end{tabular}
\end{table}

\newpage
\begin{table}[H]
\caption{Mean sum of squared deviations for each model under each data-generating condition. Bolded values indicate the best performing model based on a pairwise t-test. If multiple values are bolded, then model performances were not significantly different at 0.01 level.}
\label{tab:SSDTable}
\centering
\begin{tabular}{rrrrrr}
  \hline
 & Hierarchical & Fixed (0.5) & Full Model & Shared & Null Model \\ 
  \hline
  All Included (Prob = 0.5) & \textbf{0.0239} & 0.1023 & 0.4949 & 0.1302 & 0.5051 \\ 
   All Included (Prob = 0.1) & \textbf{0.0144} & 0.1070 & 0.9255 & \textbf{0.0225} & 0.0745 \\ 
  Independent (Prob = 0.5) & 0.0827 & \textbf{0.0773} & 0.4976 & \textbf{0.0780} & 0.5024 \\ 
  Independent (Prob = 0.1) & 0.0390 & 0.0927 & 0.9016 & \textbf{0.0287} & 0.0984 \\ 
  All Included (Prob = 1.0) & 0.0461 & 0.1250 & \textbf{0.0000} & 0.0013 & 1.0000 \\ 
  None Included (Prob = 0.0) & 0.0201 & 0.1151 & 1.0000 & \textbf{0.0003} & \textbf{0.0000} \\ 
   \hline
\end{tabular}
\end{table}

\begin{table}[H]
\caption{Mean log-posterior predictive likelihood for each model under each data-generated condition. Bolded values indicate the best performing model based on a pairwise t-test. If multiple values are bolded, then model performances were not significantly different at 0.01 level.}
\label{tab:PLTable}
\centering
\begin{tabular}{rrrrrr}
  \hline
 & Hierarchical & Fixed (0.5) & Full Model & Shared & Null Model \\ 
  \hline
  All Included (Prob = 0.5) & \textbf{-7283.03} & -7367.29 & -7494.97 & -7387.65 & -11520.08 \\ 
  All Included (Prob = 0.1) & \textbf{-7523.70} & -7620.57 & \textbf{-7863.64} & -7525.52 & -8997.26 \\ 
  Independent (Prob = 0.5) & \textbf{-7118.00} & \textbf{-7112.60} & -7321.68 & \textbf{-7114.15} & -11493.11 \\ 
  Independent (Prob = 0.1) & -7680.67 & -7754.05 & -8028.31 & \textbf{-7663.71} & -9512.73 \\ 
  All Included & \textbf{-7750.00} & -7809.16 & \textbf{-7742.74} & \textbf{-7744.13} & -12581.53 \\ 
  None Included & -7980.20 & -8070.98 & -8317.14 & -7955.02 & \textbf{-7952.85} \\ 
   \hline
\end{tabular}
\end{table}

\end{document}

\maketitle

\section{Further Details on Data Application Results}
Here we provide additional figures and discussion of the results for our application of TCGA data, expanding on our summary in Section 3.4 of the main article.  

\begin{figure}[H]
\begin{subfigure}{.5\textwidth}
\centering
\includegraphics[width=\linewidth]{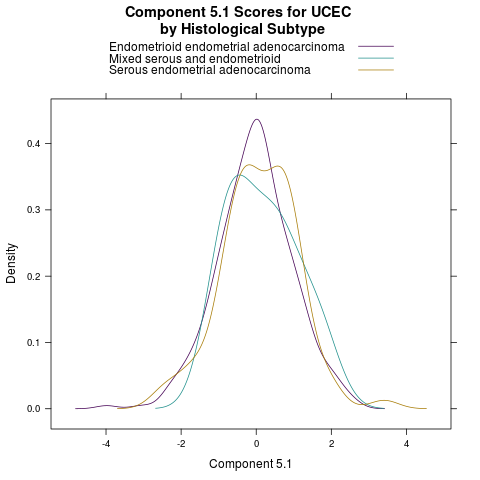}
\caption{  }
\label{fig:ucec_a}
\end{subfigure}%
\begin{subfigure}{.5\textwidth}
\centering
\includegraphics[width=\linewidth]{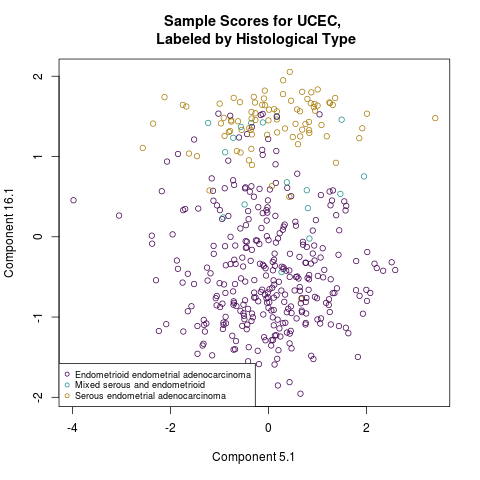}
\caption{  }
\label{fig:ucec_b}
\end{subfigure}
\begin{center}
\begin{subfigure}{.5\textwidth}
  \centering
  \includegraphics[width=\linewidth]{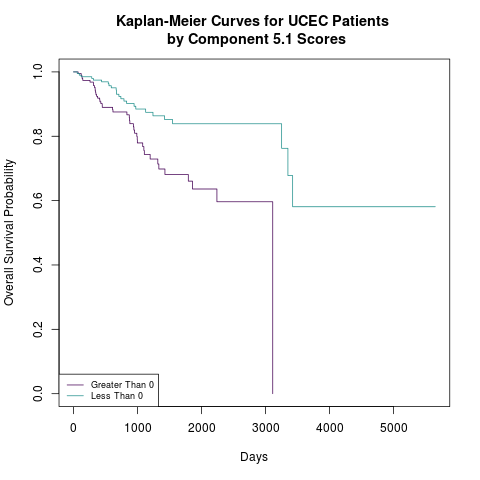}
  \caption{  }
  \label{fig:ucec_c}
\end{subfigure}
\end{center}
\caption{Figure \ref{fig:ucec_a} shows a KDE plot for subjects within each UCEC subtype along component 5.1. Component 5.1 does not appear to explain much variability in these subtypes. A similar deduction can be made from the scatterplot of subjects according to components 16.1 and 5.1 in figure \ref{fig:ucec_b}. However, this pattern of variability is marginally associated with survival, as shown in figure \ref{fig:ucec_c}, where subjects grouped by their component 5.1 scores appears to have different survival trajectories.}
\label{fig:UCEC_plot_supp}
\end{figure}

Two BIDIFAC+ components were identified as predictive of survival for UCEC subjects: 16.1 and 5.1. We observed that component 16.1 differentiated UCEC subjects according to their histological subtype: endometrioid, serous, and mixed histology; however, the link between variability explained by component 5.1 and clinical features of UCEC appeared less obvious. Figure \ref{fig:ucec_a} shows the histological subtypes of UCEC are not well differentiated by component 5.1. Figure \ref{fig:ucec_b} shows how subjects cluster along components 16.1 and 5.1 according to subtype, offering a comparison between the variation explained by either component. However, this pattern of variation characterized by component 5.1 does have a marginal association with survival. This can be observed from the Kaplan-Meier survival plot in figure \ref{fig:ucec_c}, which shows how survival outcomes differ by subjects depending on the sign of their scores for this component. 

\begin{figure}[H]
\begin{subfigure}{.5\textwidth}
\centering
\includegraphics[width=\linewidth]{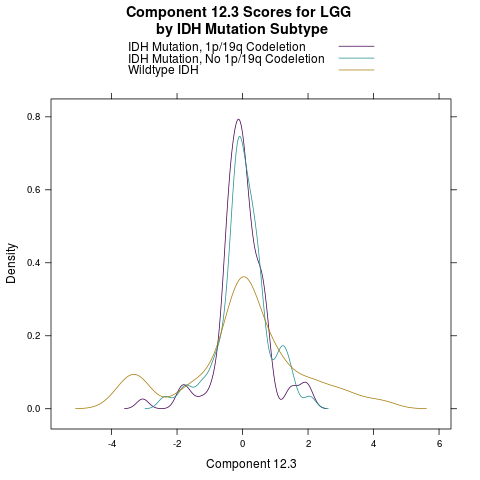}
\caption{  }
\label{fig:lgg_a}
\end{subfigure}%
\begin{subfigure}{.5\textwidth}
  \centering
  \includegraphics[width=\linewidth]{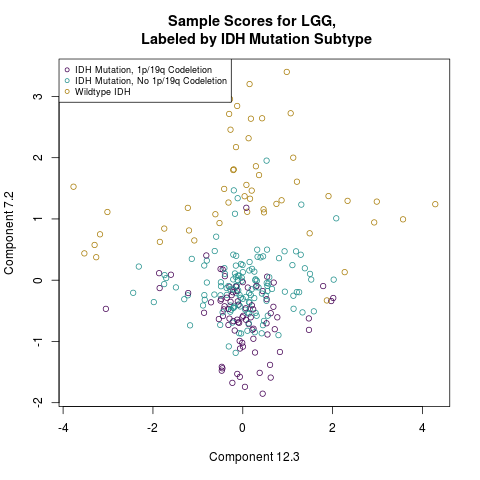}
  \caption{  }
  \label{fig:lgg_b}
\end{subfigure}
\begin{center}
\begin{subfigure}{.5\textwidth}
  \centering
  \includegraphics[width=\linewidth]{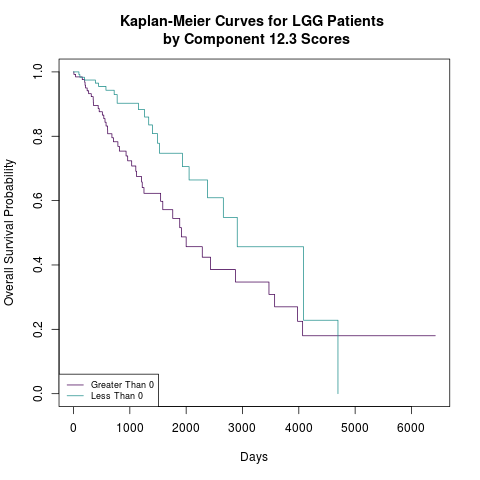}
  \caption{  }
  \label{fig:lgg_c}
\end{subfigure}
\end{center}
\caption{Figure \ref{fig:lgg_a} shows a KDE plot of subjects clustering by their IDH mutation group along component 12.3. This plot shows that this pattern explains some of the variation in IDH mutation status, but as much as component 7.2. Figure \ref{fig:lgg_b} shows how subjects cluster along both components 7.2 and 12.3. Despite a less distinct clustering pattern along component 12.3, it appears to have a marginal association with survival, as shown in figure \ref{fig:lgg_c}.}
\label{fig:LGG_plot_supp}
\end{figure}

Two BIDIFAC+ components were identified as predictive of survival in LGG, components 7.2 and 12.3. We considered the association of these components with mutation status of genes IDH1 and IDH2 and found that component 7.2 differentiates three types of IDH mutations. These mutation groups also displayed starkly different survival outcomes, with IDH wildtype mutations showing worst overall survival. Component 12.3 appeared to differentiate the IDH mutation subgroups to a lesser degree, with IDH wildtype mutations showing more variation across subjects than either of the other two groups, as shown in figure \ref{fig:lgg_a}. Figure \ref{fig:lgg_b} demonstrates how subjects cluster when plotted against component 7.2 and 12.3, allowing a comparison of how much variation each component explains. The pattern of variability characterized by component 12.3 does have a marginal association with survival, as shown by the Kaplan-Meier survival figure in figure \ref{fig:lgg_c}, which shows how survival outcomes differ by subjects depending on the sign of their component 12.3 scores.

Lastly, we considered the components identified as predictive of survival in KIRP, KIRC, and KICH. BIDIFAC+ component 11.1 was selected for both KIRP and KIRC and component 22.1 was selected for all three. In our main article, we saw KIRP subjects grouped by subtype cluster along component 11.1, with the CIMP methylator phenotype clustering very distinctly from the rest. This subgroup also showed the worst overall survival of all KIRP subtypes. Since this pattern of variability also appears to explain variation in survival outcomes in KIRC subjects, we suggested a similar CIMP phenotype may be present in KIRC, though such a subtype is not currently defined for this cancer. 

\begin{figure}[H]
\begin{subfigure}{.5\textwidth}
\centering
\includegraphics[width=\linewidth]{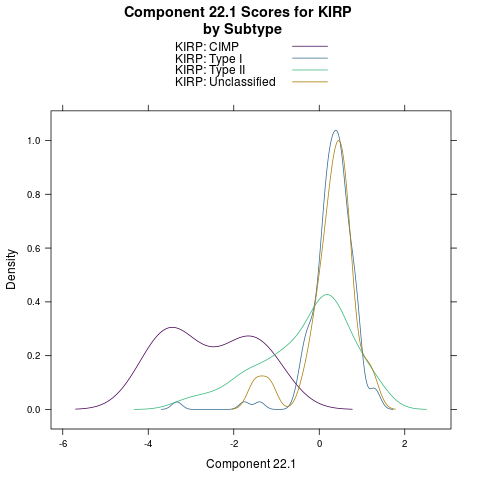}
\caption{  }
\label{fig:kidney_a}
\end{subfigure}%
\begin{subfigure}{.5\textwidth}
\centering
\includegraphics[width=\linewidth]{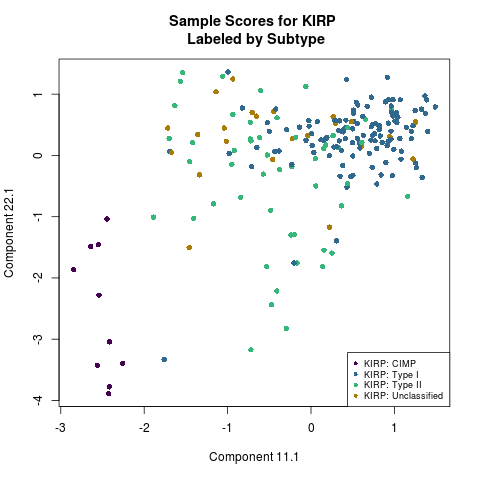}
\caption{  }
\label{fig:kidney_b}
\end{subfigure}
\begin{center}
\begin{subfigure}{.5\textwidth}
\centering
\includegraphics[width=\linewidth]{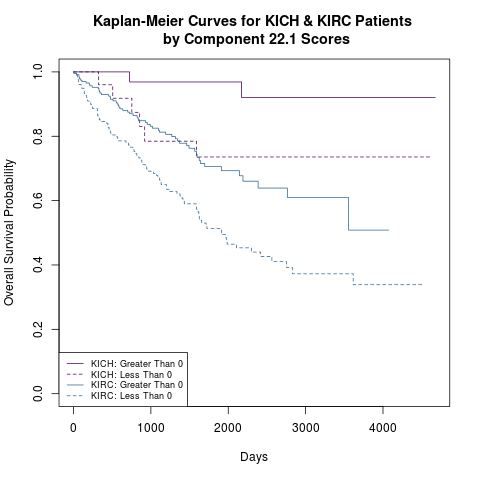}
\caption{  }
\label{fig:kidney_c}
\end{subfigure}
\end{center}
\caption{Figure \ref{fig:kidney_a} shows a KDE plot for KIRP subjects along component 22.1 clustering according to their clinical subtype. This demonstrates the CIMP subgroup clusters distinctly from the remaining groups, though not dramatically. Figure \ref{fig:kidney_b} shows how KIRP subjects cluster along components 11.1 and 22.1. Both patterns of variability appear to explain variation in the CIMP methylator phenotype in KIRP subjects to differing degrees. Despite an inability to connect this component of variation to any known clinical feature in KICH and KIRC, figure \ref{fig:kidney_c} shows that this component is marginally related to survival.}
\label{fig:Kidney_plot_supp}
\end{figure}

Component 22.1 appeared to distinguish the CIMP methylator phenotype in KIRP to a lesser degree. This can be observed in the KDE figure and scatterplot of KIRP subjects along components 11.1 and 22.1 in figures \ref{fig:kidney_a} and \ref{fig:kidney_b}. The inclusion of 22.1 for KIRC and KICH remains unclear, warranting further investigation as these are cancers that are not currently differentiated into subtypes. It seems this pattern of variation is some feature that is not directly related to histological subtypes. Nevertheless, component 22.1 is marginally associated with survival in both KICH and KIRC. This is shown in figure \ref{fig:kidney_c}, where subjects with positive and negative scores for component 22.1 have distinct survival outcomes.

\section{Validation Study}

In addition to the large-scale simulation comparison described in Section 4 of the main article, we ran two simulations to validate our Gibbs sampling algorithm and spike-and-slab model. The first simulation checked the coverage rates of credible intervals based on posteriors draws from our in-house Gibbs sampler. For each iteration of this simulation, we generated true values for each parameter from their respective priors, including inclusion indicators for the spike-and-slab component. These were then used to generate simulated data. Using these data, we generated posterior draws from our Gibbs sampler, calculated 95\% credible intervals, and checked whether the true value was indeed captured by the credible interval. The entire simulation is outlined below:
\begin{enumerate}
    \item For $iter = 1, \dots, 1000$,
    \begin{enumerate}
        \item Fix the number of clusters at a chosen value. Randomly generate sample sizes for each cluster from 50 to 500. 
        \item Generate predictors from a $\hbox{Normal}(0,1)$ to be stored in matrix $\mathbf{X}$ and generate true values for each of the model parameters from their respective priors. The true values are denoted as vectors $\beta^*$, $\tilde{\beta}^*$, $\lambda^{2*}$, $\gamma^*$ , and $\sigma^{2*}$. Based on the inclusion indicators, $\gamma^*$, the excluded $\beta^*$ will either be generated from the spike or the slab distribution. 
        \item Generate survival times from a normal distribution with mean $\mathbf{X}\beta^*$ and variance $\sigma^{2*}$. Use this distribution to generate censor times. Replace survival times with censor times if the survival time is greater than the corresponding censor time with ``not available" (NA). 
        \item Run the Gibbs sampler based on the generated data for 2000 iterations. Use a 1000 iteration burn-in and compute 95\% credible intervals.
        \item Compare the resulting credible intervals with the generated true values. Tally how many times the credible intervals contained the true parameters.
    \end{enumerate}
    \item Check that for approximately 95\% of iterations, the true value was contained in its respective credible interval.
\end{enumerate}

We set up a second simulation to assess the accuracy of the spike-and-slab component of the model. We followed an identical set up to the one above but instead of considering 95\% credible intervals, we calculated the posterior inclusion probability for each predictor by averaging the number of times a covariate was included after burn-in. If this mean was greater than 0.5, we considered that covariate as ``included," otherwise it was considered ``excluded." We then compared this posterior inclusion indicator with the true inclusion indicator and tallied the proportion of times our model correctly identified a predictor should be included. The design of this simulation is described in more detail below:

\begin{enumerate}
    \item For $i = 1, \dots, 1000$: 
    \begin{enumerate}
        \item Follow the same strategy as in the previous simulation to generate data. 
        \item Run the Gibbs sampler on the simulated data.
        \item Store the inclusion/exclusion results from the sampler after including a burn-in. If a covariate was included over 50\% of the time, the covariate was considered ``included." Otherwise, it was considered ``excluded."
        \item Compare the inclusion/exclusion results from the Gibbs sampler with the true inclusion/exclusion indicators. Tally the number of times the model correctly identified a covariate as included or excluded. 
    \end{enumerate}
    \item Confirm nominal accuracy of the model. We saw our model correctly identified predictors approximately 90\% of the time. 
\end{enumerate}

In both validation studies, we assumed there were 12 clusters. The model for each cluster contained an intercept, as well as a subset of three possible covariates. Similar to our TCGA data application, we assumed the clusters did not share identical covariate sets. We generated outcomes for each cluster from a normal distribution and censored approximately 50\% of subjects. We assigned the following priors for each of the model parameters (and these were the priors from which we generated the true values for this simulation in step 1b): 

For $i=1\dots, 12$, let $S_i = \{\ell: X_\ell \text{ is a predictor for cluster }i\}$. Assume a linear model for the $j$th response in cluster $i$:

\[
y_{ij} = \beta_{i0} + \sum_{\ell\in S_i} \beta_{i\ell} X_{ij\ell} + \epsilon_{ij}
\]

where $\epsilon_{ij} \sim N(0,\sigma^2)$. We excluded the intercept from the spike-and-slab framework and assumed $\beta_{i0} \sim \hbox{Normal}(\tilde\beta_0, \lambda^2_0)$. For the remaining predictors, $j\in S_i$:

\begin{align*}
    \beta_{ij} &\sim (1-\gamma_{ij}) \hbox{Normal}\left(0, \frac{1}{10000}\right) + \gamma_{ij} \hbox{Normal}(\tilde\beta_k, \lambda^{2}_j)
\end{align*}

Assume $\tilde\beta_0 \sim \hbox{Normal}(0, 10^2)$ and $\lambda^2_0 \sim \hbox{Inverse-Gamma}(1,1)$. Further assume $\tilde\beta_\ell \sim \hbox{Normal}(0,1)$ and $\lambda^2_\ell \sim \hbox{Inverse-Gamma}(5,1)$ for $j=1,2,3$. Lastly, we used assume the variance in the survival outcomes $\sigma^2 \sim \hbox{Inverse-Gamma}(1,1)$. Our priors and likelihood differed slightly from those used in our analysis for computational ease. \\

The coverage results for this simulation can be found in table \ref{tab:coverate_results}, which shows coverage rates for each parameter and covariate combination. The selection accuracy results can be found in table \ref{tab:selection_accuracy}. These results demonstrate nominal coverage rates which confirms the model is running properly. 

\begin{table}[H]
\centering
\begin{tabular}{|c|c|c|c|c|}
  \hline
 Parameter & Intercept $j=0$ & Covariate $j=1$ & Covariate $j=2$ & Covariate $j=3$ \\ 
  \hline
  $\tilde\beta$ & 0.944 & 0.950 & 0.961 & 0.939 \\ 
   $\lambda^2$ & 0.952 & 0.957 & 0.949 & 0.958 \\ 
   $\pi$ &  & 0.925 & 0.938 & 0.953 \\ 
  $\beta_{1.}$ & 0.951 & 0.928 &  & 0.951 \\ 
  $\beta_{2.}$ & 0.937 & 0.948 & 0.943 & 0.955 \\ 
  $\beta_{3.}$ & 0.941 &  & 0.948 &  \\ 
  $\beta_{4.}$ & 0.948 & 0.937 & 0.937 &  \\ 
  $\beta_{5.}$ & 0.947 &  &  & 0.929 \\ 
  $\beta_{6.}$ & 0.959 & 0.946 &  & 0.942 \\ 
  $\beta_{7.}$ & 0.948 &  & 0.940 &  \\ 
  $\beta_{8.}$ & 0.944 & 0.943 &  &  \\ 
  $\beta_{9.}$ & 0.947 & 0.952 & 0.954 &  \\ 
  $\beta_{10.}$ & 0.947 & 0.943 & 0.944 & 0.937 \\ 
  $\beta_{11.}$ & 0.946 & 0.948 &  &  \\ 
  $\beta_{12.}$ & 0.955 &  & 0.929 &  \\ 
   \hline
\end{tabular}
\caption{Coverage proportions for each parameter and covariate combination, showing the proportion of simulations for which the true value of each parameter was contained in its credible interval. Each row corresponds to a parameter and each column corresponds to a covariate. For $i=1,\dots, 12$, the vector $\beta_{i.}$ represents model coefficients for cluster $i$. Blank spaces in the table indicate either a group did not have a certain covariate or a fixed value. For example, the intercept was included in the model for every cluster, so there is no value for $\pi$ under the column ``Intercept."}
\label{tab:coverate_results}
\end{table}

\begin{table}[H]
\centering
\begin{tabular}{|c|c|c|c|}
  \hline
 Group & Covariate $j=1$ & Covariate $j=2$ & Covariate $j=3$ \\ 
  \hline
  1 & 0.856 &  & 0.826 \\ 
  2 & 0.919 & 0.892 & 0.891 \\ 
  3 &  & 0.885 &  \\ 
  4 & 0.926 & 0.907 &  \\ 
  5 &  &  & 0.918 \\ 
  6 & 0.912 &  & 0.924 \\ 
  7 &  & 0.904 &  \\ 
  8 & 0.912 &  &  \\ 
  9 & 0.929 & 0.899 &  \\ 
  10 & 0.868 & 0.889 & 0.875 \\ 
  11 & 0.926 &  &  \\ 
  12 &  & 0.833 &  \\ 
   \hline
\end{tabular}
\caption{Selection accuracy results for the coefficient parameters of each group. Each cell is the proportion of iterations that a particular covariate was correctly included in the model. Each row corresponds to a parameter and each column corresponds to a covariate. For $i=1,\dots, 12$, the vector $\beta_{i.}$ represents model coefficients for cluster $i$. Blank spaces in the table indicate a covariate was not available. For example, group 12 did not have covariates 1 and 3.}
\label{tab:selection_accuracy}
\end{table}